\documentclass[runningheads]{llncs}
\usepackage[justification=centering]{caption}
\usepackage{adjustbox}
\usepackage{times}
\usepackage{graphicx}
\usepackage{url}
\usepackage{amsmath}
\usepackage{amssymb}
\usepackage{listings}
\usepackage{color}
\usepackage{hyperref}
\usepackage{float}
\usepackage{caption}
\usepackage[affil-it]{authblk}
\usepackage{listings}
\usepackage{xcolor}
\usepackage{longtable}

\definecolor{codebg}{RGB}{245,245,245}
\definecolor{keyword}{RGB}{0,0,180}
\definecolor{comment}{RGB}{0,128,0}

\title{Ladder Logic Translation using Large Language Models in Industrial Automation}

\author{Oluwatosin Ogundare \inst{1,3} and
Promise Ekpo \inst{2} \and
Nathanial Wiggins \inst{1}}
\authorrunning{O. Ogundare et al.}
\institute{Industrial \& Systems Engineering, University of Houston, TX 
\email{\{oogundar,nwiggins\}@central.uh.edu}\\
\and
Computer Science, Cornell University, Ithaca, NY\\
\email{poe6@cornell.edu}
\and
Information \& Decision Sciences, California State University-San Bernardino, CA}

\begin{document}
\maketitle

\begin{abstract}
Ladder logic translation is an important problem in industrial automation because without it, it is difficult to switch Programmable Logic Controller (PLC) vendors. The prevailing translation problem highlights mismatched programming environments, incompatible ladder logic constructs, limitations in terms  of differences in the semantic expressiveness of the vendor formalisms and integrated black-box proprietary engineering tools which are exemplified in our example case; Rockwell to Siemens PLC code translation. This work presents a mathematical formulation of the problem, the detailed architecture of a solution which supports XML extraction, structural normalization, constrained generative function (LLM), and system integration via the TIA Portal Openness API as rigorously engineered pipeline for automated translation of Rockwell Ladder Programs to Siemens S7 ladder programs. Finally, we present results that show that the translations retain high semantic consistency across instruction categories. 
\end{abstract}

\section*{Introduction}
Code translation with Large Language Models (LLMs) is an important problem in enterprise software \cite{yang2024exploring}, but industrial automation remains difficult because PLC ecosystems are highly vendor-specific. For example, Rockwell Automation employs the XML-based L5X format, whereas Siemens TIA Portal exposes S7 program blocks through the TIA Openness API \cite{siemensTIAOpenness}\cite{huhtanen2024tia}. Although both platforms nominally adhere to IEC~61131-3, substantial differences remain in execution semantics, branch representations, and general instruction formalisms which limits practical interoperability despite XML standardization \cite{scott2020learning} \cite{berger2014automating}. Although, migration tools exist, they often rely on static instruction mappings that require extensive manual correction for nested branch logic and other vendor-specific constructs \cite{rockwell_plc5_logix_migration_docs}. Recent advances in LLM reasoning capabilities, together with growing adoption in industrial engineering \cite{ogundare2023industrial}\cite{ogundare2022no}, create opportunities for translating structured industrial logic through symbolic representations \cite{fakih2024llm4plc}\cite{fakih2025llm4cve}. Platform-agnostic intermediate representations have previously been explored for automation and verification tasks \cite{rodriguez2017generacion} \cite{hostetler2016process}; however, our approach performs an object-oriented reduction of serialized XML to support constrained LLM-guided translation and deterministic post-processing. Figure~1 describes the overall architecture of the translator. The remainder of this paper is organized as follows: Section~\ref{sec:theory} formalizes the translation problem as a sequence transduction over an intermediate representation. Section~\ref{sec:architecture} describes the staged pipeline, LLM translation engine, and presents quantitative evaluation results on a shift-register program. Section~\ref{sec:conclusion} concludes with future directions.

\section{Formalism of sequence transduction over IR}\label{sec:theory}
In our example case, \texttt{<Task>} elements define execution behavior and scheduling constraints, \texttt{<Program>} elements determine scan order through \texttt{<ScheduledPrograms>}, and \texttt{<Routine>} elements encode executable logic units. Therefore, we define the intermediate representation (IR) as the \(n\)-tuple, such that \(IR=(T,P(\cdot),R(\cdot),L(\cdot))\), where each component captures a distinct layer of the PLC execution hierarchy. Let \(T=\{t_i\}_{i=1}^{n}\) denote the finite set of execution tasks, with associated program mappings \(P(t_i)=\{p_{i1},p_{i2},\dots,p_{ik}\}\) and routine mappings \(R(p_{ij})=\{r_{ij1},r_{ij2},\dots,r_{ijm}\}\). Each routine containing ladder logic is then expressed as an ordered rung sequence \(L(r_{ijm})=(u_1,u_2,\dots,u_k)\).
\paragraph{\textbf{Sequence transduction formulation}} \mbox{} \\ 
We define the semantic function associated with a routine \( r \) as the composition of its constituent rung-level transformations:
\begin{equation}
\delta_r = u_k \circ u_{k-1} \circ \cdots \circ u_1
\end{equation}
Given that function composition is non-commutative in general, i.e., the ordering of transformations is essential, we know that
\begin{equation}
u_i \circ u_j \neq u_j \circ u_i
\end{equation}
It therefore follows that the mapping from an ordered sequence of instructions to its induced transformation,
\begin{equation}
(u_1, \dots, u_k) \mapsto \delta_r
\end{equation}
is inherently order-sensitive. This implies that the intermediate representation (IR) induces a sequence-dependent transformation over instruction space, i.e., a sequence transduction problem for which an LLM can be applied.
\paragraph{\textbf{Hierarchical Compositional Structure}} \mbox{}\\
Additionally, the mapping shown in (3) implies that the IR induces a structure through its hierarchy from derived higher-order compositions. Let $\bigcirc$ be a composition operator over transformation mappings. The IR induces higher-order compositions like so:
\begin{equation}
\delta_r = \bigcirc_{u \in L(r)} u, \quad
\delta_p = \bigcirc_{r \in R(p)} \delta_r, \quad
\delta_t = \bigcirc_{p \in P(t)} \delta_p
\end{equation}
Therefore, the full execution semantics of a PLC may be expressed as the nested composition:
\begin{equation}
\delta = \bigcirc_{t \in T} \delta_t
\end{equation}
This is crucial because it formalizes how complex PLC execution semantics emerge from the composition of simple instructions in a principled, order-sensitive way and fits well with sequence based ML algorithms. All symbols retain their meaning from the IR formalism. 

\section{Architecture Overview and LLM Translation Engine} \label{sec:architecture}
\begin{figure}
\centering
\includegraphics[width=0.48\textwidth]{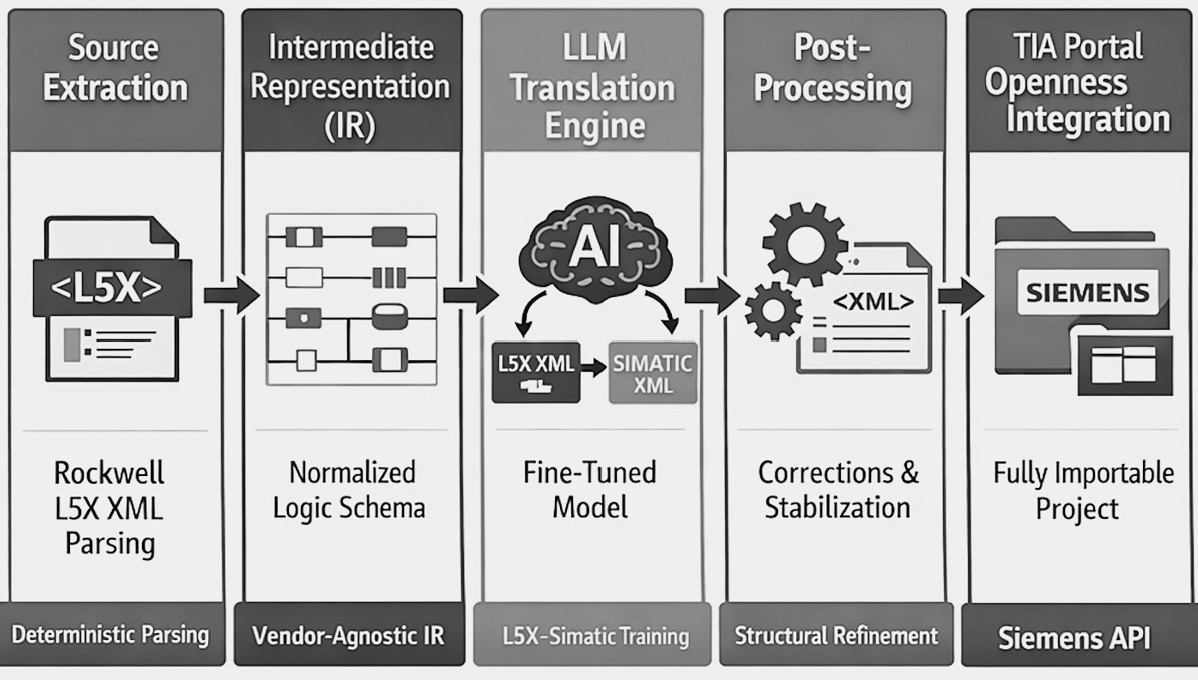}
\caption{L5X-to-SimaticXML translation pipeline}
\label{fig:architecture1}
\end{figure}

\begin{figure}[h!]
\centering
\includegraphics[width=0.57\textwidth]{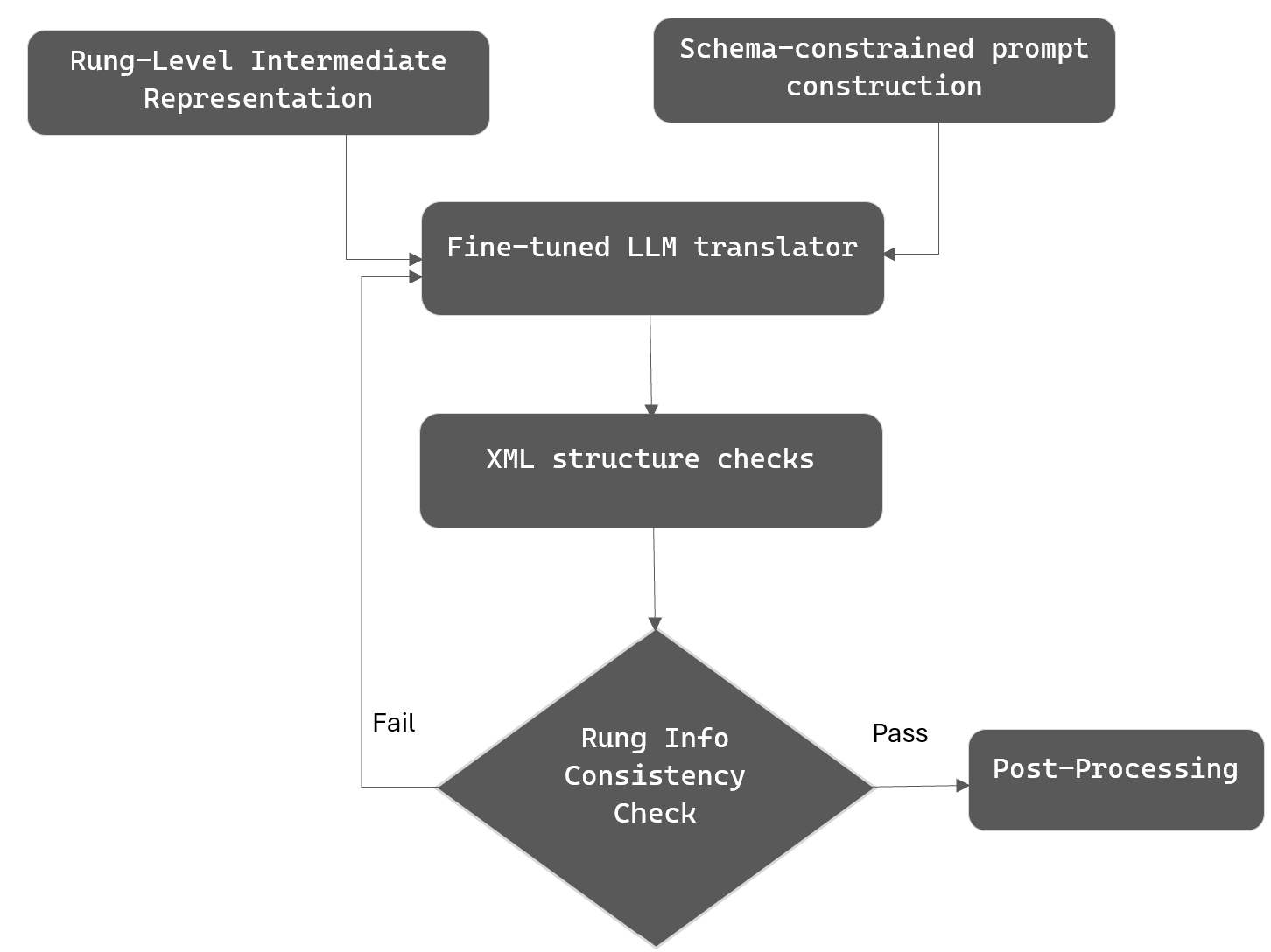}
\caption{LLM-guided rung-level translation architecture}
\label{fig:llm_translation_engine}
\end{figure}

The architecture in Figure~\ref{fig:architecture1} implements a staged pipeline that transforms Rockwell L5X ladder logic into Siemens SimaticXML importable through TIA Portal. The process begins with recovery of the L5X XML using a best-effort parser written in \texttt{C\#}, followed by construction of a deterministic object-oriented intermediate representation (IR) capturing task hierarchies, routines, rung structures, and symbolic tag references. This reduces the burden on the generative model while remaining constrained by the SimaticXML schema.
The core translation stage employs an LLM fine-tuned on aligned pairs of L5X-derived representations and SimaticXML blocks. The model is conditioned on Siemens-specific schema fragments and block patterns, enabling structural validity\cite{siemensTIAOpenness}. Deterministic post-processing then applies rule-based structural corrections for common translation failures, including ladder network reconstruction and reference normalization, before assembling a fully importable Siemens engineering project. The translation engine shown in Figure~\ref{fig:llm_translation_engine} constitutes the semantic core of the pipeline.

\begin{figure}[H]
\centering
\includegraphics[width=0.67\textwidth]{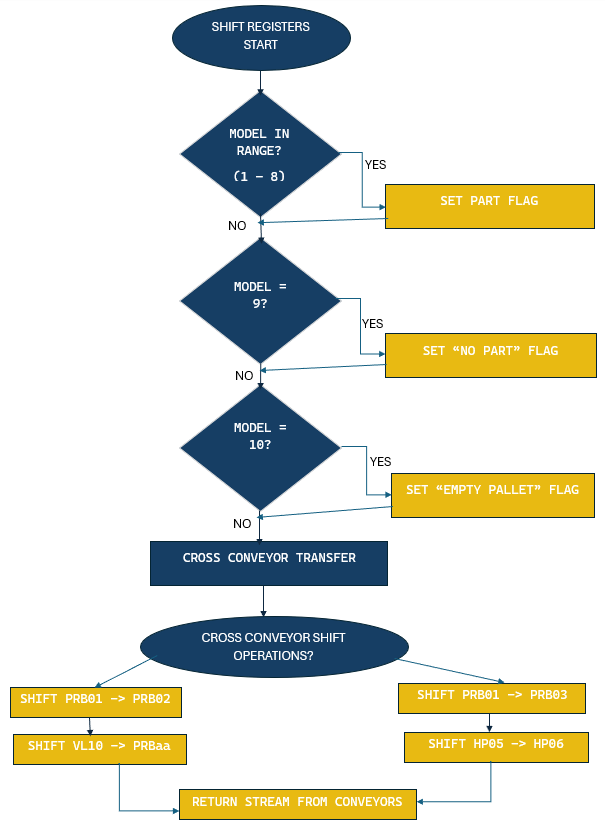}
\caption{Shift-register ladder logic in an L5X project}
\label{fig:problemset}
\end{figure}

\begin{figure}[h!]
\begin{adjustbox}{left=-5mm}
\includegraphics[width=1\textwidth]{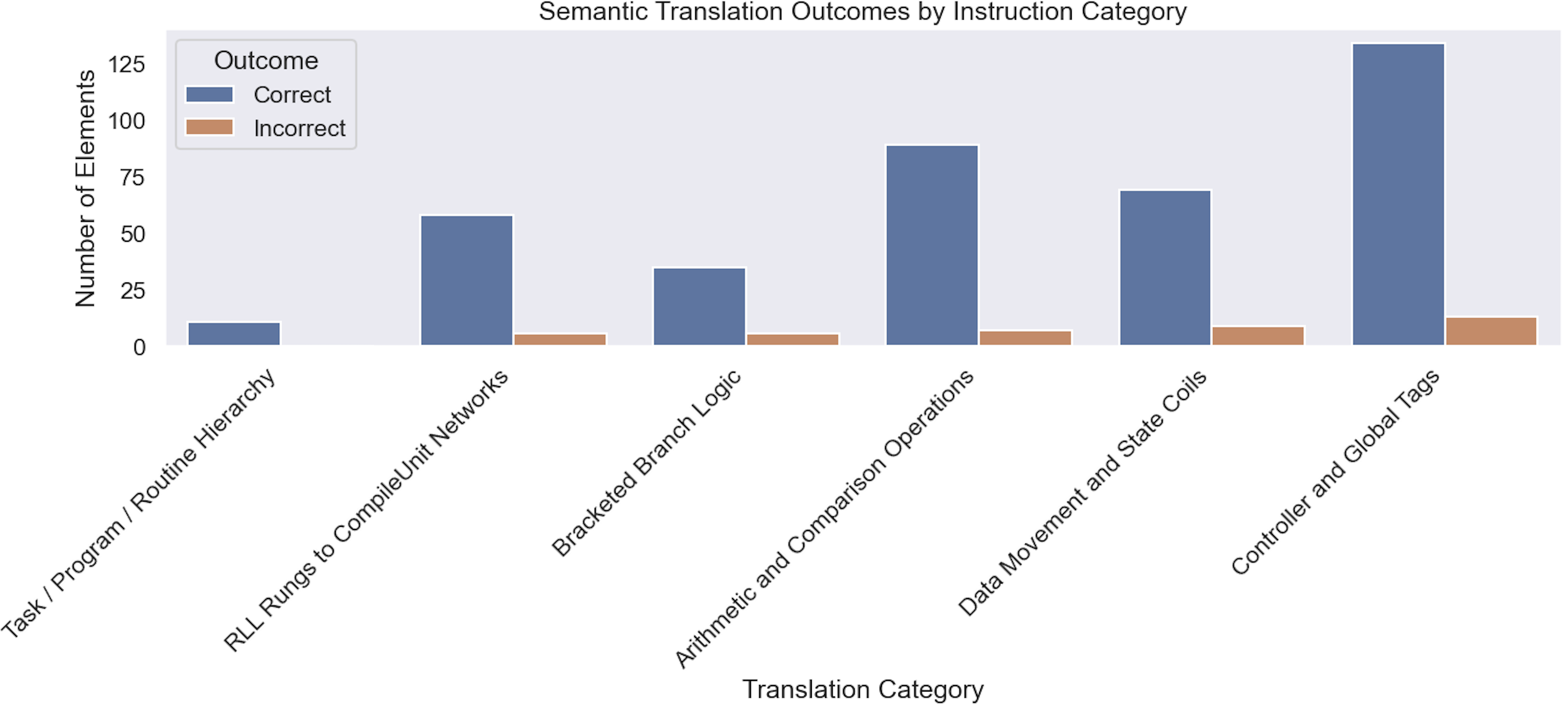}
\end{adjustbox}
\caption{Evaluation results over the categories}
\label{fig:translation_results_1}
\end{figure}

\paragraph{\textbf{Schema-constrained Prompting \& Consistency Check}} \mbox{} \\Translation prompts embed explicit structural context, including SimaticXML schema fragments, Siemens block templates (OB, FB, FC), and permissible instruction enumerations, thereby constraining generation toward structurally valid outputs. For each rung-level IR instance, the model additionally generates a natural-language behavioral explanation and derives a SimaticXML representation. A cosine similarity metric \cite{7577578} is then computed between SimaticXMLs derived from Rockwell L5X and from natural-language descriptions; outputs below a predefined threshold are rejected.
\paragraph{\textbf{Evaluation of LLM translation of Shift Registers ladder logic in an L5X project}} \mbox{}\\
Figure~\ref{fig:problemset} illustrates the shift-register control program used for evaluation. The original Allen-Bradley L5X project was translated into Siemens SimaticXML, with routines mapped into Siemens Function Blocks (FCs) and ladder rungs mapped into \texttt{CompileUnit} networks. Quantitative evaluation results are summarized in Table~\ref{tab:shift_register_translation_metrics} and Figure~\ref{fig:translation_results_1}.

\begin{table}[h!]
\centering
\begin{adjustbox}{width=\textwidth,keepaspectratio}
\begin{tabular}{|l|c|c|c|c|}
\hline
\textbf{Translation Category} &
\textbf{Total Elements} &
\textbf{Correctly Translated} &
\textbf{Partially Correct / Failed} &
\textbf{Untranslated / Failed} \\ \hline

Task / Program / Routine Hierarchy &
11 &
11 (100.0\%) &
0 (0.0\%) &
0 (0.0\%) \\ \hline

RLL Rungs $\rightarrow$ CompileUnit Networks &
64 &
58 (90.6\%) &
4 (6.3\%) &
2 (3.1\%) \\ \hline

Bracketed Branch Logic &
41 &
35 (85.4\%) &
5 (12.2\%) &
1 (2.4\%) \\ \hline

Arithmetic and Comparison Operations (EQU, GEQ, LEQ) &
96 &
89 (92.7\%) &
6 (6.3\%) &
1 (1.0\%) \\ \hline

Data Movement \& State Coils (MOV, OTL/OTU, SCoil) &
78 &
69 (88.5\%) &
6 (7.7\%) &
3 (3.8\%) \\ \hline

Controller / Program Scope Tags and Global Variables &
147 &
134 (91.2\%) &
10 (6.8\%) &
3 (2.0\%) \\ \hline

\textbf{Overall Translation} &
\textbf{437} &
\textbf{396 (90.6\%)} &
\textbf{31 (7.1\%)} &
\textbf{10 (2.3\%)} \\ \hline

\end{tabular}
\end{adjustbox}
\caption{Quantitative Evaluation of L5X to SimaticXML Translation for Shift Register Logic}
\label{tab:shift_register_translation_metrics}
\end{table}

\section{Conclusion} \label{sec:conclusion}
Our evaluation focuses on a shift register control program and is intended to validate the proposed translation architecture rather than provide comprehensive coverage of all PLC constructs. Results in Table~\ref{tab:shift_register_translation_metrics} and Figure~\ref{fig:translation_results_1} show that explicitly modeled hierarchical structures are preserved deterministically, achieving 100\% accuracy in Task / Program / Routine mappings. High accuracy in arithmetic operations (92.7\%) and tag handling (91.2\%) further suggests that constrained intermediate representations(IR) improve symbol-level translation consistency. In contrast, reduced performance in bracketed branch logic (85.4\%) and rung-to-network translation (90.6\%) highlights the primary failure mode of the system, since branch structures must often be reconstructed implicitly from L5X representations. Future work should therefore focus on extending the IR to encode branch topology and control-flow dependencies more explicitly in order to reduce ambiguity during translation.\\
\textbf{Data Availability Statement}\\
The data is summarized within the manuscript and raw files are available upon request. Please send an email to the corresponding author (Oluwatosin Ogundare) at oogundar@central.uh.edu

\bibliographystyle{IEEEtran} 
\bibliography{refs} 

\end{document}